\documentclass[varenna]{cimento}
\usepackage{graphicx}
\usepackage{amsmath}
\usepackage{multirow}
\usepackage{float} 
\usepackage{url,tabularx,amsfonts,slashed,array,cite}
\title{ADD and MLS signals in top-antitop final states}
\author{K. Mimasu and S. Moretti}
\institute{School of Physics \& Astronomy, \\ University of Southampton, \\ Southampton, SO17 1BJ, UK}
\begin{document}
	\maketitle
	\vspace{-30mm}
	\begin{abstract}
		We study top-(anti)quark pair production at the Tevatron and LHC in the context of the Minimal Length Scenario 
		(MLS) of the Arkani-Hamed, Dimopoulos and Dvali (ADD)
		model of extra dimensions (XDs). We show that sizable effects onto both the integrated and differential 
		cross section due to graviton mediation	are expected for a String scale, $M_S$, of ${\cal O}(1-10~{\rm TeV})$
		and several XDs, $\delta$, all compatible with current experimental constraints. Potential 
		limits on $M_S$ are extracted. We also highlight clear phenomenological differences between 
		a simple ADD scenario and its modification based on using the MLS as a natural regulator for divergent 
		amplitudes of virtual KK graviton exchange.
	\end{abstract}
	\vspace{-14mm}	
	\section{Introduction}
	\label{sect:intro}
	\noindent
	We present a study of the phenomenology of the Arkani-Hamed, Dimopoulos and Dvali (ADD) model~\cite{arkani1998hierarchy}
	extended with a Minimal Length Scenario (MLS) in the $t\bar{t}$ channel based on~\cite{Mimasu2011}. The focus is to
	determine the reach of the LHC to set bounds on the fundamental mass scale of the model, $M_{S}$, which is the $\mathcal{O}(TeV)$
	parameter that relates to the usual Planck scale, $M_{P}$, via the volume of some $\delta$ extra dimensions, $V_{\delta}$ as: $M_{P}^{2} = V_{\delta}M_{S}^{2+\delta}$.
	This recasts the gauge hierarchy problem in terms of compact extra dimensions in which only gravity can propagate, 
	diluting its strength relative to other forces and reconciling the observed Planck mass with the EW scale. Such models can be understood as low energy effective realisations of 
	String type theories~\cite{AntoniadisPhys.Lett.B436:257-2631998} where bulk gravitons emerge in 4 dimensions as infinite towers of Kaluza-Klein (KK) modes that 
	couple universally to the SM via the energy-momentum tensor with a strength suppressed by the Planck scale. 
	\section{ADD in the $t\bar{t}$ channel and the Minimal Length Scenario}
	\label{sect:ADD}
	\noindent
	In the $t\bar{t}$ channel, the dominant SM contribution comes from QCD in both the gluon and di-quark initial states 
	while the ADD model allows the $s$-channel exchange of a virtual KK graviton.
	The fact that the graviton is a colour singlet forbids interference with the latter while allowing it with the $t$ 
	and $u$-channel components of $gg\rightarrow t\bar{t}$. This combined with the gluon dominated, high top quark statistics expected at the LHC
	suggests that this channel may be a viable compliment to others probing this model.\\
	\noindent
	Accounting for the sum over all KK modes leads to an effective coupling of order $1/M_{S}$. The KK modes being very closely spaced, this sum can be approximated by an integral
	\begin{equation}
		D(p^{2}) = \sum_{\vec{n}}\frac{1}{p^{2}-m_{n}^{2}+i\epsilon} \Rightarrow \int^{\Lambda^{2}}\frac{\rho(m_{n})dm_{n}^{2}}{p^{2}-m_{n}^{2}+i\epsilon},
	\end{equation}
	The density function, $\rho$, accounts for the degeneracy in KK levels. The tree-level 
	amplitude for the mediation of a virtual KK graviton in the $s$-channel gains an effective coupling factor, $\lambda_{\rm{eff}}$, from this integral: 
	\begin{equation}\label{eq:addint}
		\lambda_{\rm{eff}}\sim D(s)\kappa^{2} \sim \frac{F(\hat{s},\delta)}{M_{S}^{4}}P\left[I\left(\tfrac{M_{S}}{\sqrt{\hat{s}}}\right)\right];\quad I\left(\tfrac{M_{S}}{\sqrt{\hat{s}}}\right)=\int^{M_{S}/\sqrt{\hat{s}}}_{0}dy\frac{y^{\delta-1}}{1-y^{2}},
	\end{equation}
	where $\hat{s}$ is the partonic centre of mass energy while $P$ denotes the principal part 
	of the integral. The integral is divergent and necessitates a hard cutoff, usually associated with $M_{S}$.\\
	The MLS, defined as the incorporation of a minimum length $l_{S}$ into the ADD setup, provides an alternative 
	mechanism for regularising these divergences. This is done by modifying the relationship between 
	momentum $p$ and wave vector $k$ such that, whilst particle momentum can become arbitrarily high, the wave vector 
	is bounded and reaches a plateau at $1/l_{S}$. A simple parametrisation that captures the essence of a 
	MLS is that of the Unruh relations~\cite{PhysRevD.51.2827}:
	\begin{equation}
	\begin{split}
		k(p)=\frac{1}{l_{S}}\text{tanh}^{1/\gamma}\left[\left(\frac{p}{M_{S}}\right)^{\gamma}\right];\qquad
		\omega(E)=\frac{1}{l_{S}}\text{tanh}^{1/\gamma}\left[\left(\frac{E}{M_{S}}\right)^{\gamma}\right],
	\end{split}
	\end{equation} 
	where $\gamma$ is henceforth assumed to be 1 for simplicity\footnote{Different choices of $\gamma$ would simply correspond to alternative ways
	of reaching such a plateau, while the asymptotic behaviour would remain the same. The limit $\gamma \rightarrow \infty$ tends toward recovering 
	a hard cutoff at $M_{S}$.}. As discussed in~\cite{Hossenfelder:2003jz}, this smoothly cuts off the integral by modulating the density function with 
	a factor $\partial \omega/\partial E$ such that:
	\begin{equation}\label{eq:mlsint}
		I^{\prime}=\int^{\infty}_{0}dy\frac{y^{\delta-1}}{1-y^{2}}\text{sech}^{2}\left(\frac{\sqrt{\hat{s}}}{M_{S}}y\right)
	\end{equation}
	as well as by modifying the phase space measure. The integral becomes finite and can be evaluated numerically.\\
	\section{Results}
\begin{figure}[h!]
		\centering
		\includegraphics[width=0.32\linewidth]{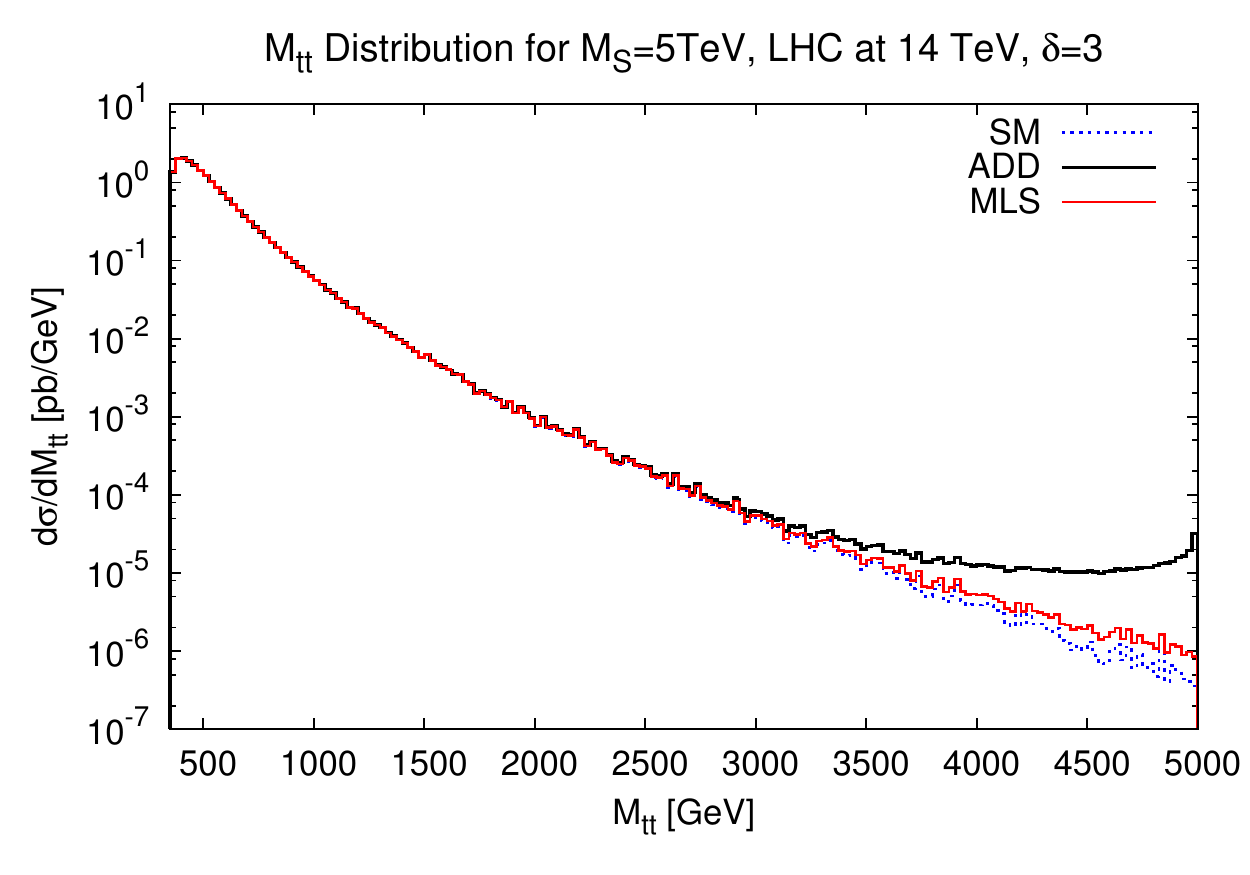}
		\includegraphics[width=0.32\linewidth]{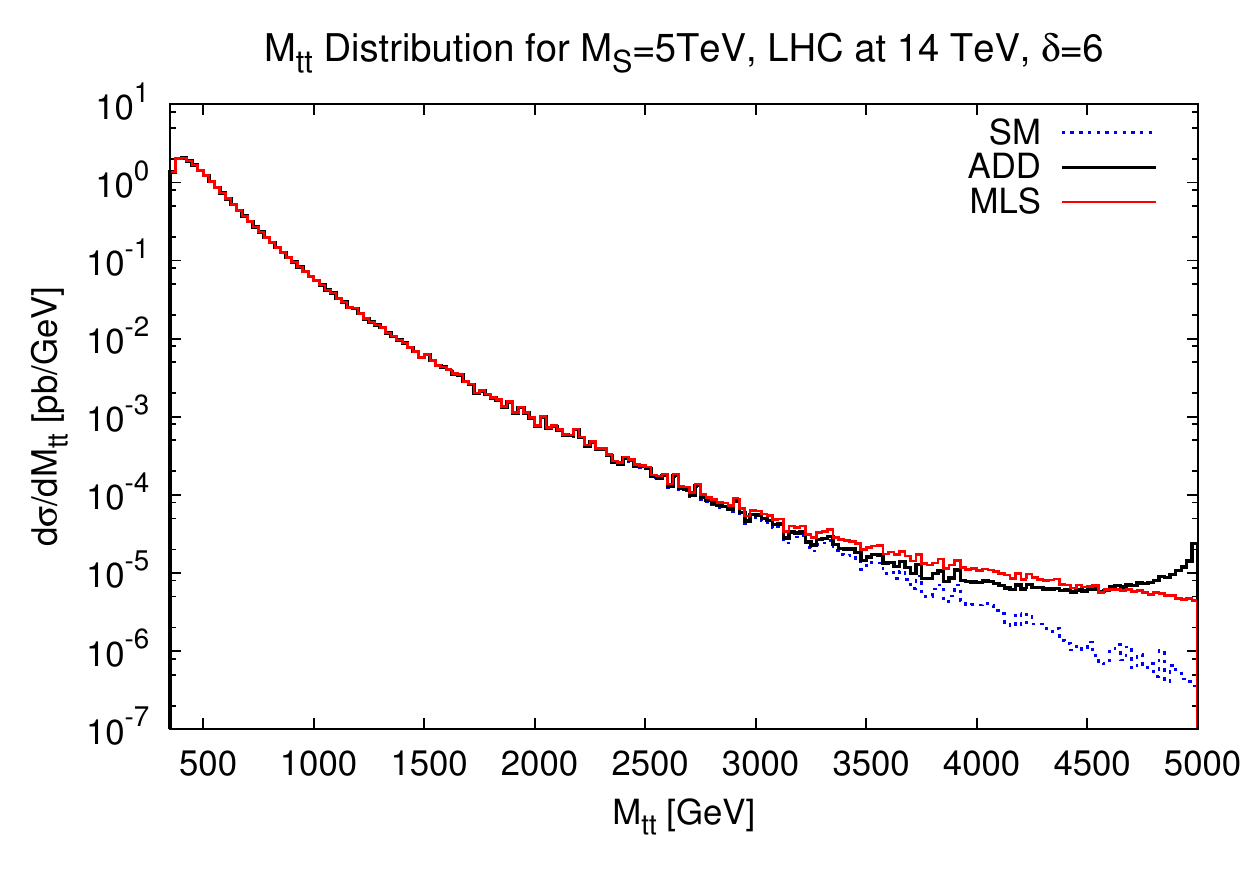}
		\includegraphics[width=0.32\linewidth]{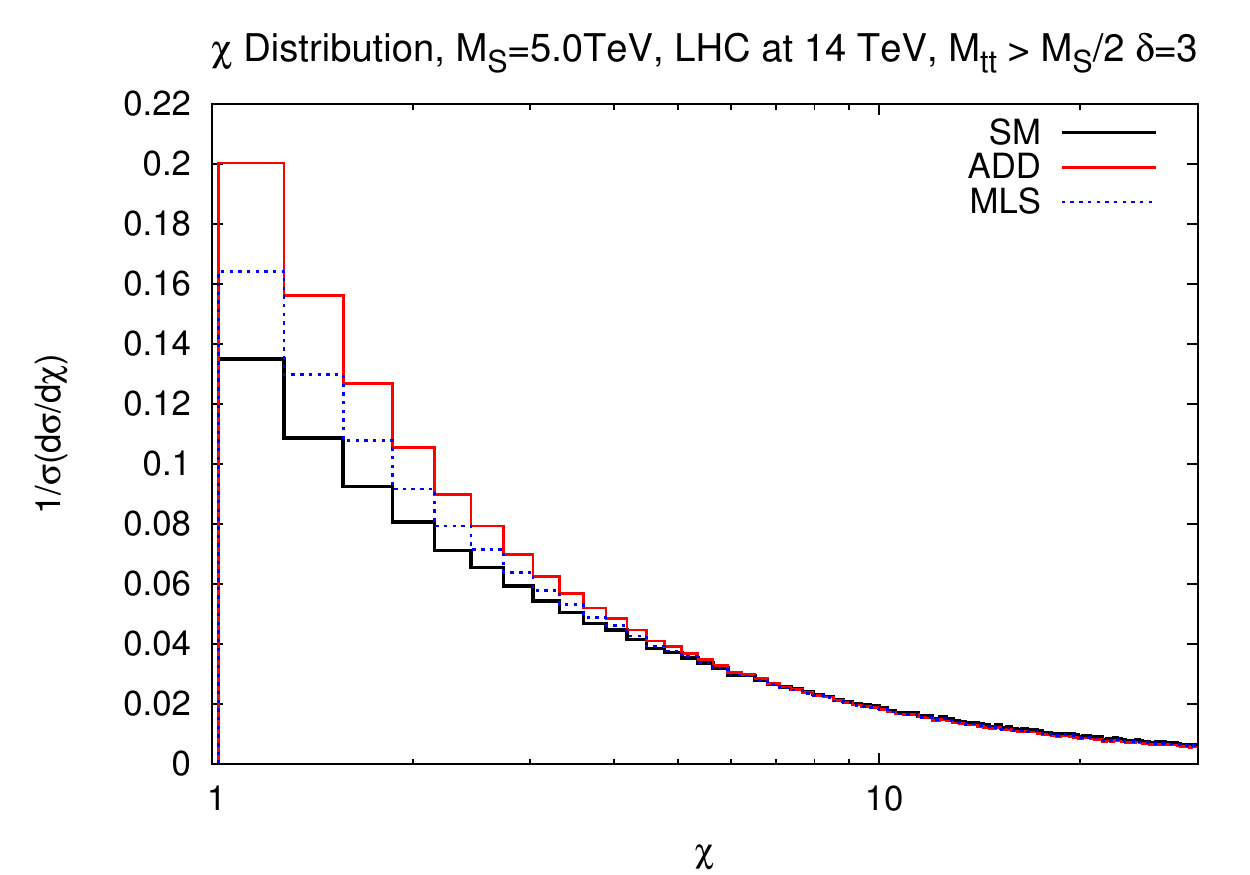}\\
		\includegraphics[width=0.49\linewidth]{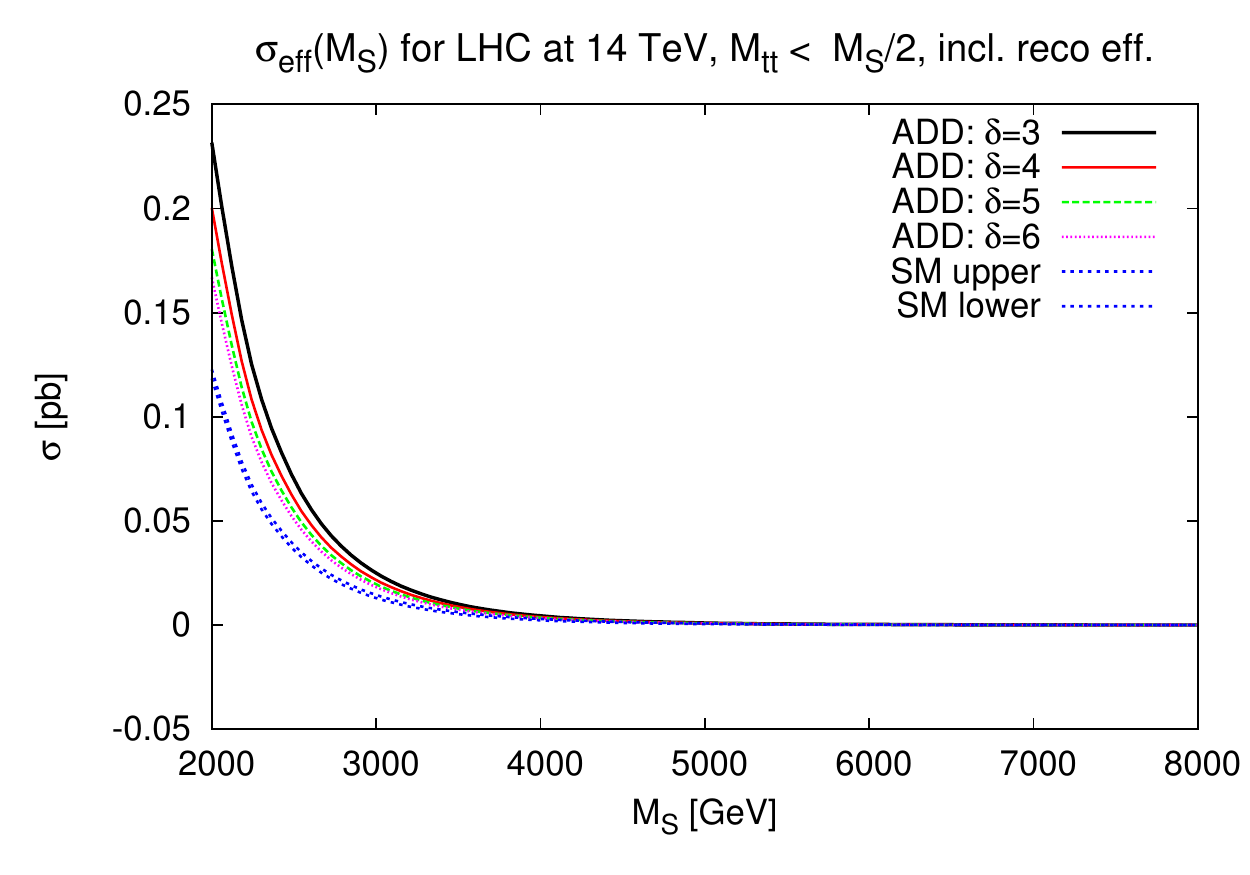}
		\includegraphics[width=0.49\linewidth]{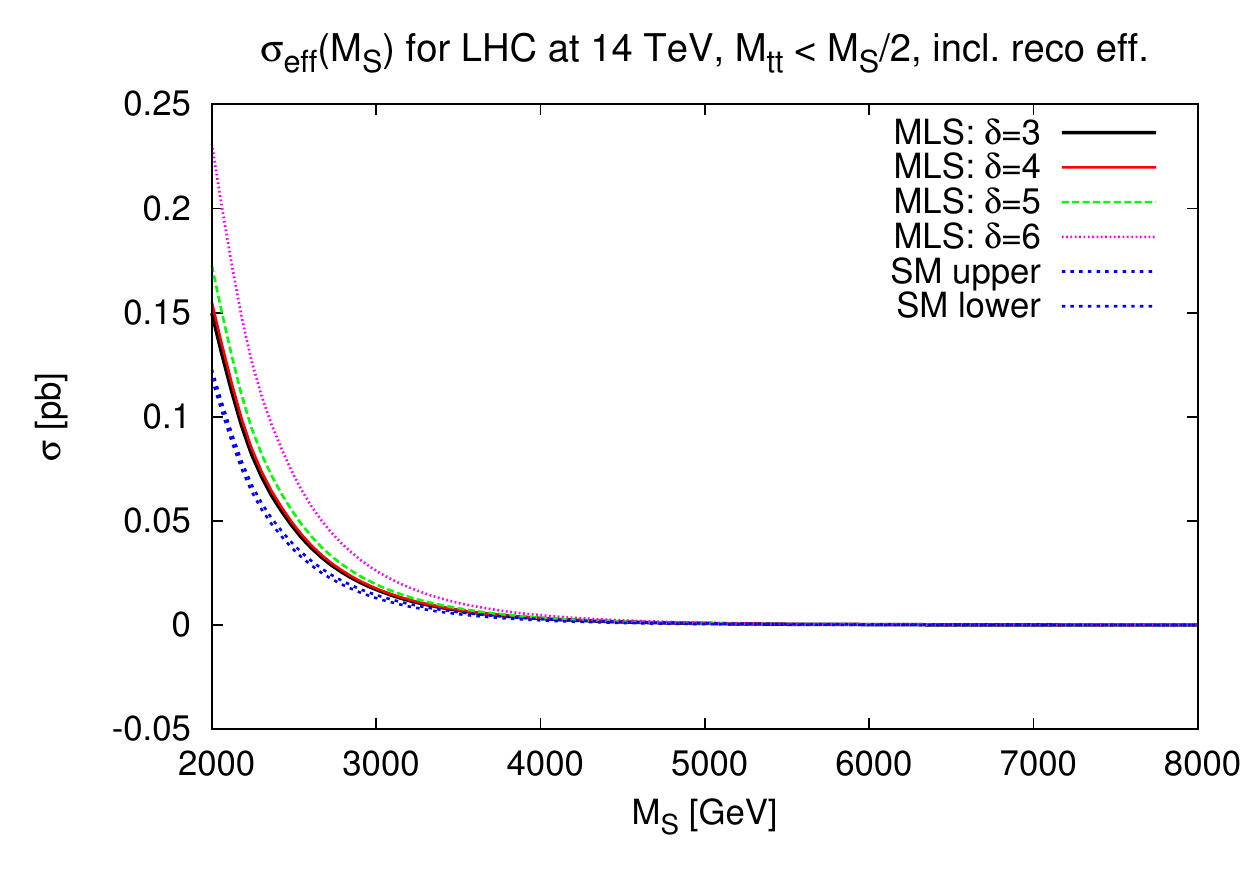}
		\vspace{-3mm}
		\caption{\emph{Top from left to right}: $t\bar{t}$ invariant mass distributions comparing the SM to ADD with and without MLS for $M_{S}$=5 TeV and $\delta$=3 
		(\emph{first}) and 6 (\emph{second}) at the LHC with $\sqrt{s}$=14 TeV; $\chi$ distributions for $M_{S}$=5 teV and $\delta=$ 3 normalised to 1 with $M_{tt}<M_{S}/2$.\newline
		\emph{Bottom}:Total cross section for $pp\rightarrow t\bar{t}$ at the LHC with $\sqrt s=$~14 TeV as a function of $M_S$ for $\delta~=~3-6$ XDs in ADD and ADD-MLS
		separately.Also plotted are the SM 95\% confidence level (CL) upper and lower bounds for 100 fb$^{-1}$ of integrated luminosity. 
		Cuts were enforced as described in the text and an estimate of 4\% for the $t\bar{t}$ reconstruction efficiency folded in.}\label{fig:plots}
	\end{figure}

\noindent	
	The ADD contributions to this process were calculated according to the conventions of~\cite{Han:1998sg} and based on the 
	matrix elements published in~\cite{Mathews:1998kf}, folding in the modifications from the MLS. 
	See~\cite{Mimasu2011} for a more detailed description of the calculation and numerical simulation for which we present a few distributions 
	and reach plots scanning the $(M_{S},\delta)$ parameter space in Fig.~\ref{fig:plots}.\\
	\noindent
	Invariant mass ($M_{tt}$) distributions give some insight into the behaviour 
	near the cutoff, $M_{S}$, where one would expect the effective model to start breaking down. 
	This is reflected in Fig.~\ref{fig:plots} where the ADD cross section appears in fact to diverge near $M_{S}$.
	The absence of this in the MLS shows that the mass integral suppression and phase space factors tame $\lambda_{\rm{eff}}$ at high invariant mass. 
	We argue that a valid comparison can occur only if one does not integrate up to $M_{S}$ but up to $\sim0.8M_{S}$, 
	where an unnatural enhancement of the pure ADD cross section begins, avoiding an over-estimation of the ADD contribution. 
	Overall, order 100\% effects could be found at energies around the cutoff. Guided by these distributions, we found that a tracking invariant 
	mass cut of $M_{tt}<M_{S}/2$ gave a satisfactory signal-to-background ratio. The $\chi = \exp(\vert y_{1}-y_{2}\vert)$ distributions also show significant deviations 
	and motivated a cut of $\chi<4$. Indeed, this particular observable may be one of the best choices for 
	directly setting bounds on signals induced by virtual gravitons, as applied in~\cite{Franceschini2011}.
	\noindent
	Using this information, a series of plots were produced of the total $pp\rightarrow t\bar{t}$ cross section as a function of $M_{S}$ implementing the aforementioned cuts
	and compared to 95\% CL limits on the SM cross section as shown in the lower half of Fig.~\ref{fig:plots}. These were used to determine the potential reach of the given collider benchmark 
	to set bounds on $M_{S}$, summarised in table~\ref{tab:limits}. 
	\begin{table}[h!]
	\begin{center}
	\begin{tabular}{cccc}
	\hline
	Collider & Luminosity &$M_S$ in ADD& $M_S$ in ADD-MLS\\
	\hline
	LHC at 14  TeV& 100 fb$^{-1}$ &5.4 TeV&5.1 TeV\\
	LHC at 7  TeV& 5 fb$^{-1}$ &2.2 TeV&2 TeV\\
	Tevatron at 2 TeV & 4 fb$^{-1}$ & $<$1 TeV&$<$0.8 TeV\\
	\hline
	\end{tabular}\caption{The 95\% CL limits on the reach to set bounds on $M_S$ in ADD and ADD-MLS at the three collider benchmarks considered. The efficiency
	of reconstructing the $t\bar t$ final states in all possible decay channels is included and
	cuts on $M_{tt}$ and $\chi$ were enforced, as described in the text. }
	\label{tab:limits}
	\end{center}
	\end{table}
	\vspace{-10mm}
	\section{Conclusions}\noindent
	The LHC at 14 TeV has a reach well beyond current experimental bounds around 5 TeV while the LHC at 7 TeV has a more limited but still potentially relevant reach given the fact that
	about twice as much will be collected than assumed at the time of the study. The Tevatron is barely able to make any observations due to the low energy and the dominance of the $q\bar{q}$.
	In particular, we also highlighted the danger of overestimating the ADD cross section due to the unreliable behaviour near the cutoff and how this is cured when implementing a smooth 
	cutoff procedure for the KK graviton sum as brought about by the MLS. Consequently the MLS lowers the reach of colliders to set bounds on $M_{S}$. 
	It was also observed that the MLS contributions increased and surpassed pure ADD with increasing $\delta$ as they integrate 
	over the whole KK spectrum. Overall, this channel can prove useful in probing such XD models as a complement to other, perhaps `cleaner' channels such as DY, 
	not only in measurements of the total cross section but also by considering differential distributions such as the $M_{tt}$ and $\chi$.\\
	The authors thank the NExT Institute for partial financial support.
	\\
	\vspace{-12mm}
	\bibliography{references}
\end{document}